\documentclass[a4paper]{acmart}

\usepackage{blindtext}
\usepackage{hyperref}
\hypersetup{
    colorlinks=true,
    linkcolor=blue,
    filecolor=magenta,      
    urlcolor=cyan,
    pdftitle={Overleaf Example},
    pdfpagemode=FullScreen,
    }
\setcopyright{acmlicensed}
\copyrightyear{2024}
\acmYear{2024}
\acmDOI{XXXXXXX.XXXXXXX}

\begin{document}
\title{MaSkel: A Model for Human Whole-body X-rays Generation from Human Masking Images}

\author{Yingjie Xi}
\email{yxi@bournemouth.ac.uk}
\affiliation{%
  \institution{National Centre for Computer Animation of Bournemouth University}
  \country{UK}
}

\author{Boyuan Cheng}
\affiliation{%
  \institution{National Centre for Computer Animation of Bournemouth University}
  \country{UK}
}
\email{bcheng@bournemouth.ac.uk}

\author{Jingyao Cai}
\affiliation{%
  \institution{National Centre for Computer Animation of Bournemouth University}
  \country{UK}
}
\email{s5604009@bournemouth.ac.uk}

\author{Jian Jun Zhang}
\affiliation{%
 \institution{National Centre for Computer Animation of Bournemouth University}
 \country{UK}}
\email{jzhang@bournemouth.ac.uk}

\author{Xiaosong Yang}
\affiliation{%
  \institution{National Centre for Computer Animation of Bournemouth University}
  \country{UK}}
\email{xyang@bournemouth.ac.uk}


\begin{abstract}
The human whole-body X-rays could offer a valuable reference for various applications, including medical diagnostics, digital animation modeling, and ergonomic design. The traditional method of obtaining X-ray information requires the use of CT (Computed Tomography) scan machines, which emit potentially harmful radiation. Thus it faces a significant limitation for realistic applications because it lacks adaptability and safety. In our work, We proposed a new method to directly generate the 2D human whole-body X-rays from the human masking images. The predicted images will be similar to the real ones with the same image style and anatomic structure. We employed a data-driven strategy. By leveraging advanced generative techniques, our model MaSkel(Masking image to Skeleton X-rays) could generate a high-quality X-ray image from a human masking image without the need for invasive and harmful radiation exposure, which not only provides a new path to generate highly anatomic and customized data but also reduces health risks. To our knowledge, our model MaSkel is the first work for predicting whole-body X-rays. In this paper, we did two parts of the work. The first one is to solve the data limitation problem, the diffusion-based techniques are utilized to make a data augmentation, which provides two synthetic datasets for preliminary pretraining. Then we designed a two-stage training strategy to train MaSkel. At last, we make qualitative and quantitative evaluations of the generated X-rays. In addition, we invite some professional doctors to assess our predicted data. These evaluations demonstrate the MaSkel's superior ability to generate anatomic X-rays from human masking images. The related code and links of the dataset are available at \url{https://github.com/2022yingjie/MaSkel}.
\end{abstract}




\maketitle

\section{Introduction}
In recent years, there are many advanced works in the human field\cite{zhang2021pymaf, zhang2023pymaf, tian2023recovering, xiu2022icon, xiu2023econ} have witnessed significant developments, ranging from nearly accurate 2D human surface extraction to the reconstruction of corresponding 3D human surface meshes. However, most of these works have primarily concentrated on the human surface, whose practicability and impact are somewhat limited within the medical and animation field. One of the primary reasons for this limitation is that these field's greater emphasis on understanding human inner structures\cite{teixeira2018generating,keller2022osso,keller2023skin,li2021piano,li2022nimble}, as these could provide more critical reference for medical teaching or realistic animation character modeling. Unfortunately, few models have been designed to generate human internal organs or skeletons.

The human inner skeleton serves as a critical foundation for medical students and doctors to deepen their understanding of human anatomy. It also provides essential details such as arm length and shoulder width for animators modeling human-like characters. Traditionally, the X-rays are used to provide information about the human skeleton. However, they typically require CT scans, which emit potentially harmful radiation. Besides, the scan is time-consuming and expensive, making it impractical for daily use. In response to these challenges, this study proposes a novel method for generating human X-rays from 2D masking images with the unique shape and posture of the subject. Masking images are semantic representations of the human surface, which are simplified to two-pixel values: pixels representing the human figure are white, while background pixels are black. Our model, named MaSkel, establishes a correlation between a masking image and its corresponding X-ray. Specifically, MaSkel receives a human masking image and generates a customized X-ray image. It is crucial to note that these generated X-rays are approximations and cannot replace actual medical X-rays for diagnosis purposes. The primary goal of our research is to produce pseudo-X-ray images that are anatomically representative and consistent with the pose observed in the human surface image. This innovative approach could be used to advance the digital modeling of human anatomy, providing a valuable reference for applications in medical diagnostics, digital animation, and ergonomic design.

Accurately generating human skeleton X-rays with rich details presents a formidable challenge, primarily due to the strict constraints in the image feature space. To achieve a high level of accuracy, the generated images must maintain structural integrity, local fidelity, and an overall visual effect that closely resembles the real X-rays. Additionally, the skeletal pose of X-rays must be consistent with the human pose of masking images provided. To overcome these challenges and improve the generation quality, we design a two-stage training strategy. In the first stage, we focus on training an encoder based on masking strategy\cite{he2022masked,dosovitskiy2020image,pathak2016context}, that is capable of compressing X-rays into a latent representation with minimal loss of information. In the second stage, we train the other encoder, which accepts masking images and outputs a latent representation that is as close to the corresponding X-ray's latent representation from the first-stage encoder as possible. Then, we employ a Vector Quantized Variational AutoEncoder(VQ-VAE) strategy\cite{van2017neural,esser2021taming} to generate human X-rays that exhibit high fidelity and realism. The use of it not only facilitates the creation of detailed anatomical images but also ensures that the generated posture aligns with the masking images. This two-stage approach allows for the generation of pseudo-X-ray images that meet the requirements of structural and visual accuracy.

\begin{figure}[h]
  \centering
  \includegraphics[width=\linewidth]{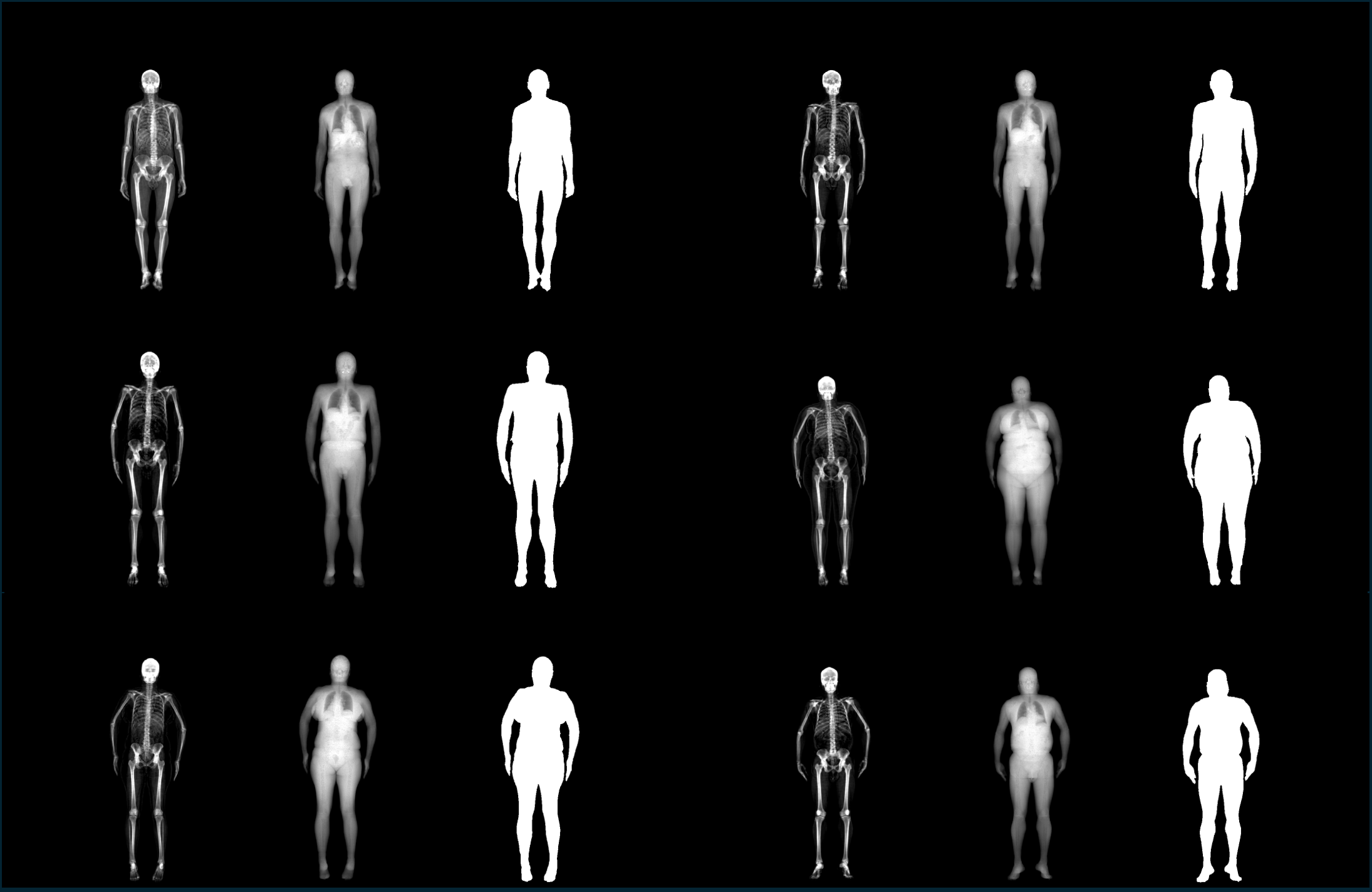}
  \caption{Paired of X-ray images(left), human soft-tissue images(middle), and masking images(right).}
  \label{img1}
\end{figure}

One of the primary obstacles in our work is the limitation of data, there are few high-quality human whole-body X-ray datasets, which significantly hinders the related research. To address this challenge, our research utilizes paired dual-energy X-ray absorptiometry (DXA)(Hereinafter referred to as X-rays) data from the UK Biobank\cite{sudlow2015uk}. This dataset consists of two types of images obtained by combining X-rays at two different energy levels: one highlighting soft tissue and the other revealing the structure and shape of bones, as illustrated in Figure\ref{img1}. Leveraging this data, we could easily get masking images from the soft tissue images by using semantic segmentation, and build paired data including masking and X-rays. Because the size of datasets is important for deep learning tasks, based on the UK Biobank database, we also build synthetic human X-ray datasets with 10000 images for our following study.    

\textbf{The pivotal contributions of this paper are as follows}:

\textbf{\emph{1.As we know, this is the first work to propose a model to generate human X-ray images from human masking images.}}

\textbf{\emph{2.We have built two synthetic human X-rays datasets in $64\times64$ and $256\times256$ resolution respectively, each comprises $10,000$ images.}}

\textbf{\emph{3.We design a two-stage training strategy for the generation of customized X-ray images from masking images.}}

\textbf{\emph{4.Through the application of similarity metrics and a user study conducted with medical professionals, we demonstrate the efficacy of our method in generating well-structured and anatomic human X-rays.}}

\section{Related Work}
We review works on the human internal structure generation first and then recap the related techniques we use in this work. 

\subsection{Generating Human Skeleton}
The most work for generating human inner structure is to get the human skeleton. There have been several studies focusing on skull generation from head surface data. For instance, a statistical shape model (SSM)\cite{lugadilu2017statistical} is proposed for the skull. It aligned corresponding 3D surfaces and performed principal component analysis (PCA) to generate the SSM model using a mean shape and modes of variation from the mean. However, their approach shares similarities with the Basel Face Model\cite{paysan20093d} (BFM), which has a limited basis for generating new skull models. Besides, PCA struggles to extract deep features of the human skull, particularly due to its intricate and complex topological structure. Coarse PCA may result in missing important information.

The skeleton method\cite{keller2022osso} was developed to fit a collection of bones to X-ray images by incorporating various constraints and subsequently learned a parametric model of skeleton shape variation. They initially employed silhouette-based techniques \cite{zuffi2018lions} to extract the 3D skeleton from different patients\cite{zuffi2015stitched,zuffi20173d}. Furthermore, several works have been proposed for creating human skeleton models\cite{gilles2010creating,saito2015computational,kadlevcek2016reconstructing,zhu2015adaptable,wang2020deep}. The recent work SKEL proposes a parametric skeleton model that can be driven to simulate different pose and shape from similar to SMPL\cite{loper2023smpl,pavlakos2019expressive}. The central issue with most current methods for 3D internal structure generation is their lack of detailed anatomical information, leading to highly simplified representations. Specifically, in the context of 3D skeleton prediction, projects like OSSO\cite{keller2022osso} and SEKL\cite{keller2023skin} depend on parameters estimated from the SMPL (Skinned Multi-Person Linear) model. This reliance introduces significant error accumulation, as the SMPL model's estimations of body shape often diverge substantially from an individual's actual physical form.

To address this gap, our model focuses on 2D images to generate human whole-body X-rays, which are closer to reality and personality. In this way, this work can be used to provide well-structured human skeleton information as 2D constraints for future 3D mesh generation.

\subsection{Diffusion-based Image Generation}
Before the emergence of diffusion models\cite{ho2020denoising,yang2023diffusion,rombach2022high,yu2023video}, GANs\cite{kingma2013auto,asperti2021survey} or VAEs\cite{goodfellow2020generative,iglesias2023survey,wang2018high,razavi2019generating,jang2019recurrent} methods are all tried to generate images from sampled white noise. Building upon this concept, diffusion models take inspiration from thermodynamics and propose a different approach. Mathematically, this process can be modeled as a Markov process, which consists of two processes: The forward process and the reverse process. In the forward process, the diffusion model gradually adds Gaussian noise to an image until it becomes completely degraded into a noisy image, and in the reverse process, it removes the noise step by step to eventually generate a clear image. If a deep neural network model can effectively model this reversed process, it would be thought to have already understood the image distribution. Adding noise and removing noise, destroying the original distribution, and learning data distribution, form the fundamental idea behind diffusion models. The forward process gradually adds Gaussian noise onto the input data $x_{0}$ in $T$ steps, producing a series of noisy samples $\{x_{t}\}_{1:T}$. Noise is scheduled by $\{\beta_{t}\}_{1:T}$.
\begin{equation}
q(x_{t}|x_{t-1}) = N(x_{t};\sqrt{1-\beta_{t}}x_{t-1},\beta_{t}I)
\end{equation}

\begin{equation}
q(x_{1:T}|x_{0}) = \prod_{t=1}^{T}{q(x_{t}|x_{t-1})}
\end{equation}
As $T\rightarrow\infty$, $x_{t}$ gradually becomes a Gaussian distribution. The diffusion model randomly samples a Gaussian noise and runs diffusion in reverse to generate a new image. In the reverse process, a neural network $p_{\theta}$ is trained to match the posterior $q(x_{t-1}|x_{t})$, $p_{\theta}$ is learned by optimizing the ELBO in VAE:

\begin{equation}
p_{\theta}(x_{t-1}|x_{t}) = N(x_{t-1};\mu_{\theta}(x_{t},t),\Sigma_{\theta}(x_{t},t))
\end{equation}

There are many researches\cite{ho2022classifier,nichol2021glide,zhang2023adding,ruiz2023dreambooth,hertz2022prompt} on unconditional and conditional image generation. In this work, we only use the unconditional image technique, because our training dataset just has one class. We plan to enlarge the original dataset from the UK Biobank to generate synthetic data, more data could improve the stability and generalization of the model. With limited computing resources, at first, we have to generate low-resolution images. \textbf{We first build a $10000$ synthetic human X-ray images dataset with the size of $64\times64$}.

\begin{figure*}[t]
  \centering
  \includegraphics[width=\linewidth]{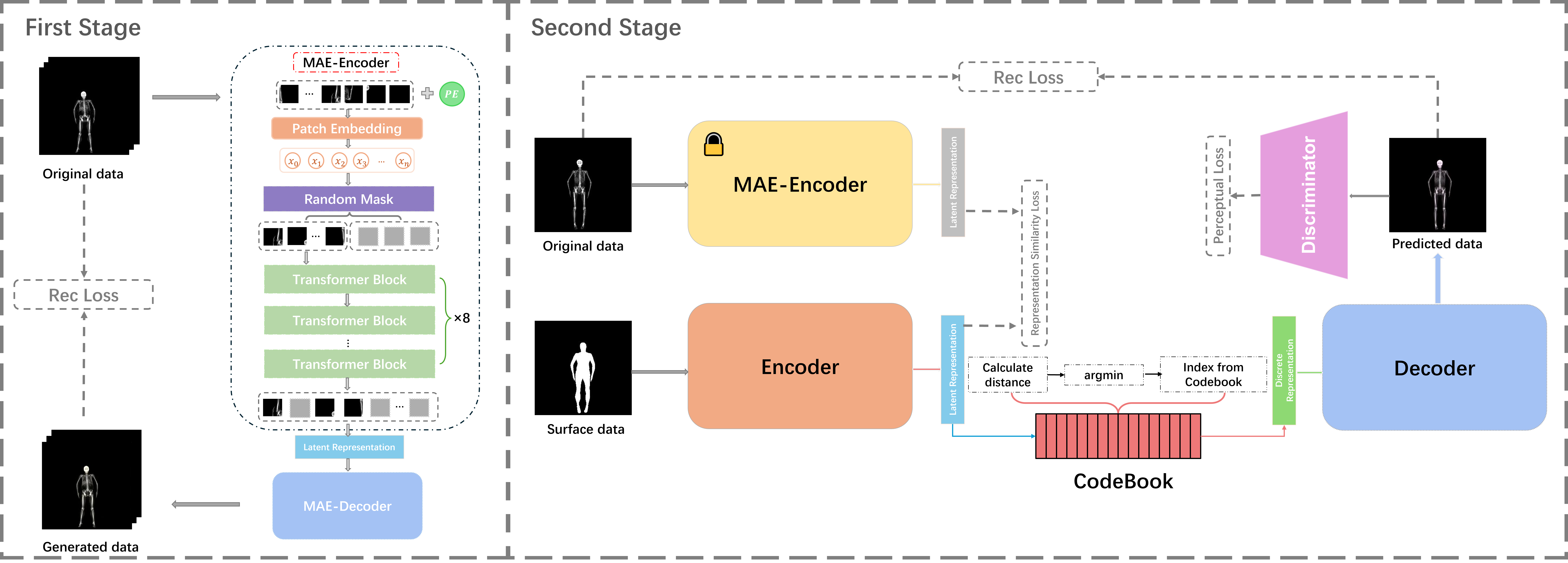}
  \caption{The left part is the details of the first stage. The right part describes the second stage structure. In the first stage, we utilize the MAE strategy to train an encoder for compressing X-rays. In the second stage, the MAE encoder accepts X-rays and outputs their latent representation as feature Ground Truth, then the other encoder maps the masking images to similar feature space. The masking feature will be fed into the decoder to generate X-ray images by using the VQ-VAE method.}
  \label{img2}
\end{figure*}

\subsection{Diffusion-based Image Super Resolution}
Image super-resolution is a fundamental problem\cite{park2023perception,fei2023generative,yao2023local} in the computer vision field, it aims at recovering high-resolution(HR) image given the low-resolution(LR) one. This problem is severely ill-posed due to the complexity and unknown nature of degradation models in real-world scenarios. For diffusion-based super-resolution tasks, one common approach\cite{saharia2022image, rombach2022high} involves inserting the LR image into the input of the current diffusion model and retraining the model from scratch on the training data for SR. Some methods\cite{choi2021ilvr,yue2022difface,wang2023exploiting} use an unconditional pre-trained diffusion model as a prior and modify its reverse path to generate the expected HR image. Unfortunately, both strategies inherit the Markov chain underlying DDPM, which can be inefficient in inference, often taking hundreds or even thousands of sampling steps. Although some acceleration techniques\cite{nichol2021improved,lu2022dpm} have been developed to compress the sampling steps in inference, they inevitably lead to a significant drop in performance, resulting in over-smooth results. 

Recently, the work ResShift\cite{yue2024efficient} model the residual between the LR and HR images as $e_{0}, e_{0}=y_{0}-x_{0}$, in which $y_{0}$ and $x_{0}$ represent the LR and HR images respectively. The core idea is to transit from $x_{0}$ to $y_{0}$ by gradually shifting residual $e_{0}$ through a Markov chain with length $T$, the formula is described as follows:

\begin{equation}
q(x_{t}|x_{t-1},y_{0}) = N(x_{t};x_{t-1}+\alpha_{t}e_{0},\kappa^{2}\alpha_{t}I)
\end{equation}

The coefficient $\kappa$ is a hyper-parameter controlling the noise variance. Similarly, the $x_{t}$ could be inferred directly from $x_{0}$.

\begin{equation}
q(x_{t}|x_{0},y_{0}) = N(x_{t};x_{0}+\eta_{t}e_{0},\kappa^{2}\eta_{t}I)
\end{equation}

And the Reverse Process aims to estimate the posterior distribution $p(x_{0}|y_{0})$ via the following formulation:
\begin{equation}
p(x_{0}|y_{0}) = \int{p(x_{T}|y_{0})\prod_{t=1}^{T}p_{\theta}(x_{t-1}|x_{t},y_{0})dx_{1:T}}
\end{equation}

ResShift could significantly accelerate the sampling speed because it chooses to shift the residual from the LR image to the HR image, which is similar to deterministic sampling. A larger skip-steps strategy could be used to speed the reverse process, and the results generated still keep high quality.

Based on the LR synthetic human X-ray dataset of size $64/times64$, we utilize ResShift to generate high-resolution X-ray images. \textbf{We build a size of $256\times256$ human X-ray skeleton dataset having 10000 images, which not only keep the whole structure of related low-resolution images but also reconstruct reasonable and anatomic structures similar to real x-rays. All images have different shapes and details to keep the diversity of the dataset}.

\subsection{MAE and VQ-VAEs}
The Masked Autoencoder(MAE)\cite{he2022masked} presents a simple technique focusing on reconstructing an original signal from its partially observed data. Different from traditional Autoencoder(AE) frameworks, MAE proposes an asymmetric design. This design enables the encoder to process only the partial, observed signal(without mask tokens) while employing a lightweight decoder to reconstruct the entire signal by integrating the latent representation with mask tokens. For the Masking strategy, MAE follows the ViT framework. It divides an image into regular non-overlapping patches. Then it samples a subset of patches and masks the remaining ones, the sampling follows a uniform distribution. The masking ratio largely eliminates redundancy to create a task that cannot be easily solved by extrapolation from visible neighboring patches. This approach significantly brings the difficulties to model in learning the right latent representation, but on the other side, this strategy could force the model to better understand the data essential features, which sets MAE to achieve high-quality data compression and reconstruction.

Vector Quantized Variational AutoEncoder (VQ-VAE) is an innovative approach in the generative field, primarily focusing on learning complex data distributions. VQ-VAE stands out by combining the principles of variational autoencoders(VAEs) with vector quantization techniques. The core idea behind VQ-VAE is to map input data into a discrete latent space rather than the continuous latent space used by traditional VAEs. This mapping is achieved through a vector quantization process, where each input data point is associated with the closest vector in a predefined set of vectors (codebook). The discrete nature of the latent space allows VQ-VAE to effectively model data distributions and generate high-quality samples. Therefore, VQ-VAE has been successfully applied in various domains, including image and speech generation, demonstrating its versatility and effectiveness in capturing intricate patterns in data.

In summary, in this paper, diffusion-based techniques are used to generate and build synthetic human X-ray datasets for research, and MAE and VQ-VAEs strategies are used for our model training based on these synthetic datasets.

\section{Method}
In this section, we introduce our method for generating human X-ray images from masking images. The MaSkel model is structured into three main components: a Vision Transformer (ViT-based Encoder)\cite{dosovitskiy2020image}, a codebook akin to that used in VQ-VAE, and a residual Network\cite{he2016deep} (ResNet)-based Decoder. Additionally, we employ a two-stage training process for our model, which is illustrated in Figure \ref{img2}. To streamline the experiment, the aforementioned neural network modules are reused across different stages.

\subsection{Two-stage Training}
In the first stage, we focus on training an encoder to capture a high-quality latent representation by using the Masked AutoEncoder (MAE) strategy. This involves randomly masking portions of the input data, enabling the encoder to better understand X-ray images. The decoder reconstructs the data from this latent representation at the pixel level. The process starts by dividing the input into patches and converting them into embeddings with added positional information. After randomly masking parts of these patches, only the unmasked tokens are processed by the ViT-based encoder. Subsequently, the masked tokens are added back into their original position to build complete a patch sequence, and they are fed into the decoder to reconstruct data.

In this part, we select the MSE loss function as shown in the following, $p_{i,j}$ is the original pixel value, and $\hat{p_{i,j}}$ means the reconstructed one.
\begin{equation}
L_{1}=\sum_{i=0}^{m}\sum_{j=0}^{n}(||p_{i,j} - \hat{p_{i,j}}||^{2})
\end{equation}

We train the encoder on our synthetic X-ray images built before. After the first-stage training, we drop the decoder and only use the encoder for the second-stage training. 

The second stage focuses on predicting X-rays by using our MaSkel model. We employ two ViT-based encoders with identical structures: the pre-trained encoder A(EA) from the first stage and an untrained encoder, encoder B(EB), alongside a decoder and a codebook. The EA accepts the X-rays to output latent representations as feature Ground Truth and the human masking images are fed into EB to get latent feature. We use the MSE loss to minimize the distance between these two representations to make EB map the masking images into the latent space of X-rays as similarly as possible. Then the VQ-VAE strategy is utilized, which employs a codebook to convert continuous latent representations into discrete codes. Each output vector $z_{e}(x)$ from the encoder is matched with its nearest neighbor $z_{q}(x)$ in the codebook, then the $z_{e}(x)$ is replaced by $z_{q}(x)$ as the input of the decoder. At last, we adopt the reconstruction loss to keep the generation quality and a perceptual loss\cite{johnson2016perceptual} to keep the generation style. Therefore, our training process is governed by three losses, which are written as follows:

\begin{equation}
L_{2}= \alpha_{1} \cdot (L_{1} + L_{q})+ 
        \alpha_{2} \cdot L_{G}(X,\hat{X}) + \alpha_{3} \cdot ||V_{e}-\hat{V}||^{2}
\end{equation}

\begin{equation}
L_{q}= ||sg[z_{e}(x)]-z_{q}||_{2}^{2} + \beta||z_{e}(x)-sg[z_{q}]||_{2}^{2}
\end{equation}

$L_{G}(X,\hat{X})$ is the perceptual loss between the realistic one and the reconstructed one respectively, $V_{e}$ is the representation of EA, and $V$ belongs to EB, the $\alpha_{1,2,3}$ is the coefficients to weight different losses. In loss $l_{q}$, $sg[\cdot]$ means gradient stop operation. The $l_{q}$ maximizes the similarity between the vector of the codebook and the vector from the encoder.

\subsection{The Structure of MaSkel}
In this section, we introduce the detailed network structure of MaSkel. The input data $X$ with the size $256\times256$ and we divide it into a sequence of 16x16 patches. 

The ViT encoder is constructed with a stack of transformer encoder blocks, each including a multi-head attention layer and a feed-forward network. In our design, the network incorporates 8 layers of transformer encoder blocks. In the design of our decoder architecture, we adopt a multi-layer design to enhance the model's ability to generate high-fidelity outputs from compressed latent representations. The core of our architecture comprises a five-layer upsampling mechanism, meticulously engineered to progressively increase the spatial resolution of the feature maps. Each upsampling layer employs a combination of linear interpolation and convolutional layers, a method proven to effectively mitigate the common pitfalls of checkerboard artifacts, thereby ensuring a smoother gradient flow and finer detail preservation in the generated output. Besides, our decoder's architecture adds several self-attention layers. This design enables the model to capture long-range dependencies across different regions of the input feature map, which is critical for generating complex patterns and textures. The codebook is composed of learned vectors throughout the training phase, with each vector signifying a discrete code. The encoder embeddings are mapped to the closest codebook vector to quantize the continuous input space into discrete codes. This mechanism offers a concise and efficient representation space, facilitating the effective compression and generation of data. 

In general, the discrete coding may lose information density and reduce the decoding space. However the distribution range of human X-rays tends to be more constrained compared to the natural images, they have less diversity and more consistency in structure. Therefore we think this discrete coding is suitable for this situation with a relatively small solution space, and it could make the training process efficient at the same time.

\section{Experiments}
In the experiments part, we have run four experiments, which include unconditional image generation, image super-resolution, image self-supervised training, and the final image generation task. The first two experiments are for data augmentation, in which we build two synthetic X-ray image datasets for the following task. Then we quantitatively evaluate reconstructed results in the first training stage, to evaluate its reconstruction quality. At last, we quantitatively and qualitatively evaluate how accurate the results we predict from the human masking images, and visualize them to see their authenticity. In addition, we invite some professional doctors to assess our predicted results, which could show our work's authenticity and reliability from the medical side.

\begin{figure}[h]
  \centering
  \includegraphics[width=\linewidth]{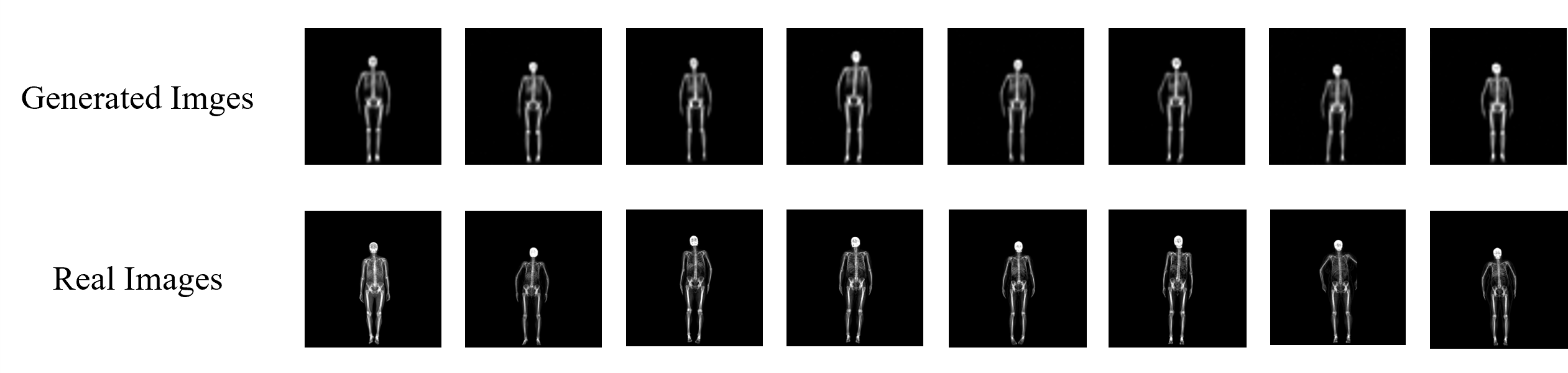}
  \caption{$64\times64$ resolution X-rays generation. This upper line is generated data and the lower line is real images. It can be seen that the generated images have the same style and structure as the real image.}
  \label{img3}
\end{figure}

\subsection{Images Augmentation}
We first generate the synthetic images with the resolution of $64\times64$ by an unconditional diffusion process. The generated results are shown in Figure\ref{img3}. 

\begin{figure}[h]
  \centering
  \includegraphics[width=1.0\linewidth]{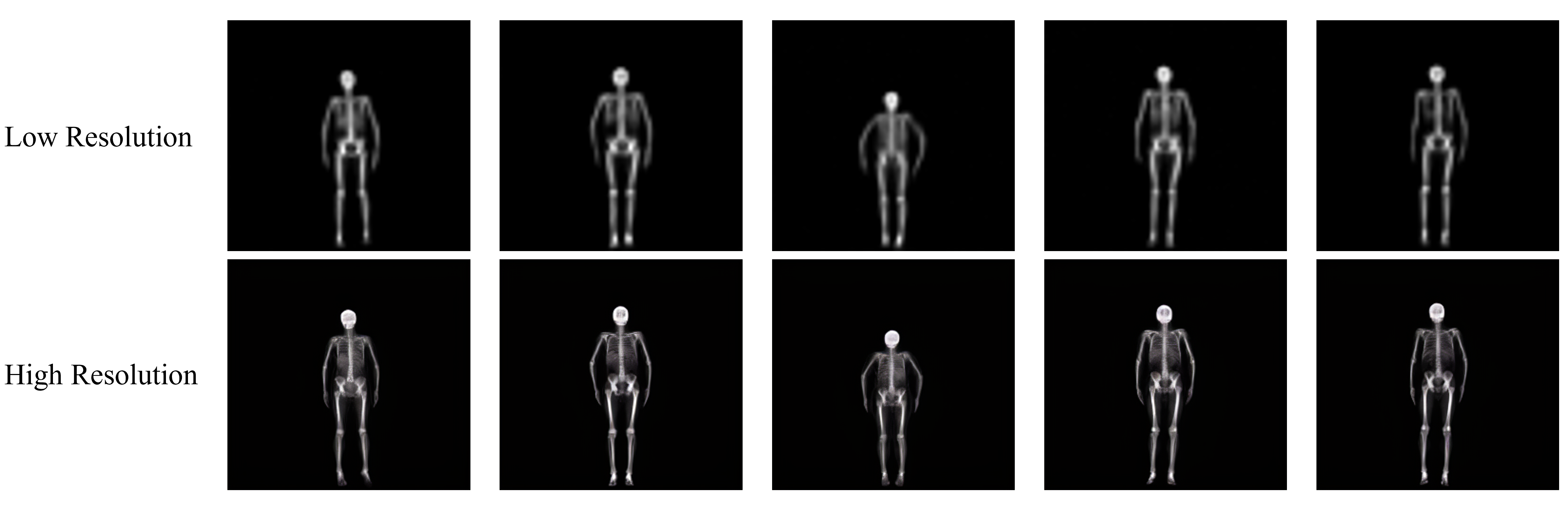}
  \caption{$256\times256$ resolution X-rays generation after super-resolution augmentation. This upper line is low-resolution data and the lower line is data with higher resolution. It can be seen that the high-resolution images have the same join and bone connection as the low-resolution ones but with far more details.}
  \label{img4}
\end{figure}

Then we use diffusion-based super-resolution techniques to increase the resolution of the image as shown in Figure\ref{img4}. The first line is low-resolution data and the second one is high-resolution data. The high-resolution image inherits the basic joint and bone connection shapes of the low-resolution image while providing more detail, with the outline, internal structure, and joint space of the bone.

\subsection{The First-stage Training}
\begin{figure}[h]
  \centering
  \includegraphics[width=1.0\linewidth]{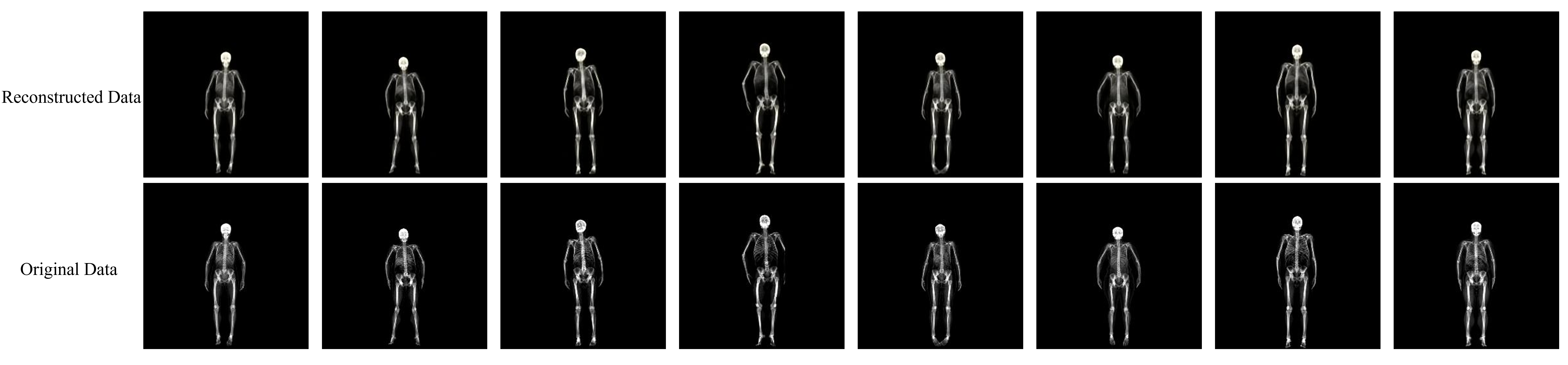}
  \caption{The first line is the reconstructed data, and the second line shows the original data. It can be seen that the reconstructed images are highly close to the original ones, which means the encoder could enable a low-loss compression.}
  \label{img5}
\end{figure}

We use the MAE method to train an encoder on our synthetic data. The reconstructed results are shown in Figure\ref{img5}, which shows that the MAE could enable a low-loss compression and well-structured X-ray image reconstruction. We evaluate the quality of reconstructed data and original data by using metrics of PSNR (Peak Signal-to-Noise Ratio), SSIM (Structural Similarity Index), and LPIPS (Learned Perceptual Image Patch Similarity) as shown in Table\ref{tabel1}. 

PSNR is a widely used metric for measuring the quality of reconstructed images compared to the original ones. It is a log scale measure of the peak error between the original and a compressed or reconstructed image.PSNR is expressed in decibels (dB), and a higher PSNR value indicates that the reconstruction is of higher quality.  SSIM is a metric that measures the similarity between two images, it considers changes in structural information, luminance, and contrast, rather than just the pixel-wise errors. The SSIM index can range from -1 to 1, where 1 indicates perfect similarity. LPIPS is a metric to assess the perceptual similarity between two images by using deep learning features to better align with human judgment. It uses features extracted from pre-trained deep networks and calculates the distance between these features to estimate perceptual similarity. A lower value of LPIPS indicates that the two images are more similar, Otherwise, the difference is greater.

\begin{figure*}[t]
  \centering
  \includegraphics[width=\linewidth]{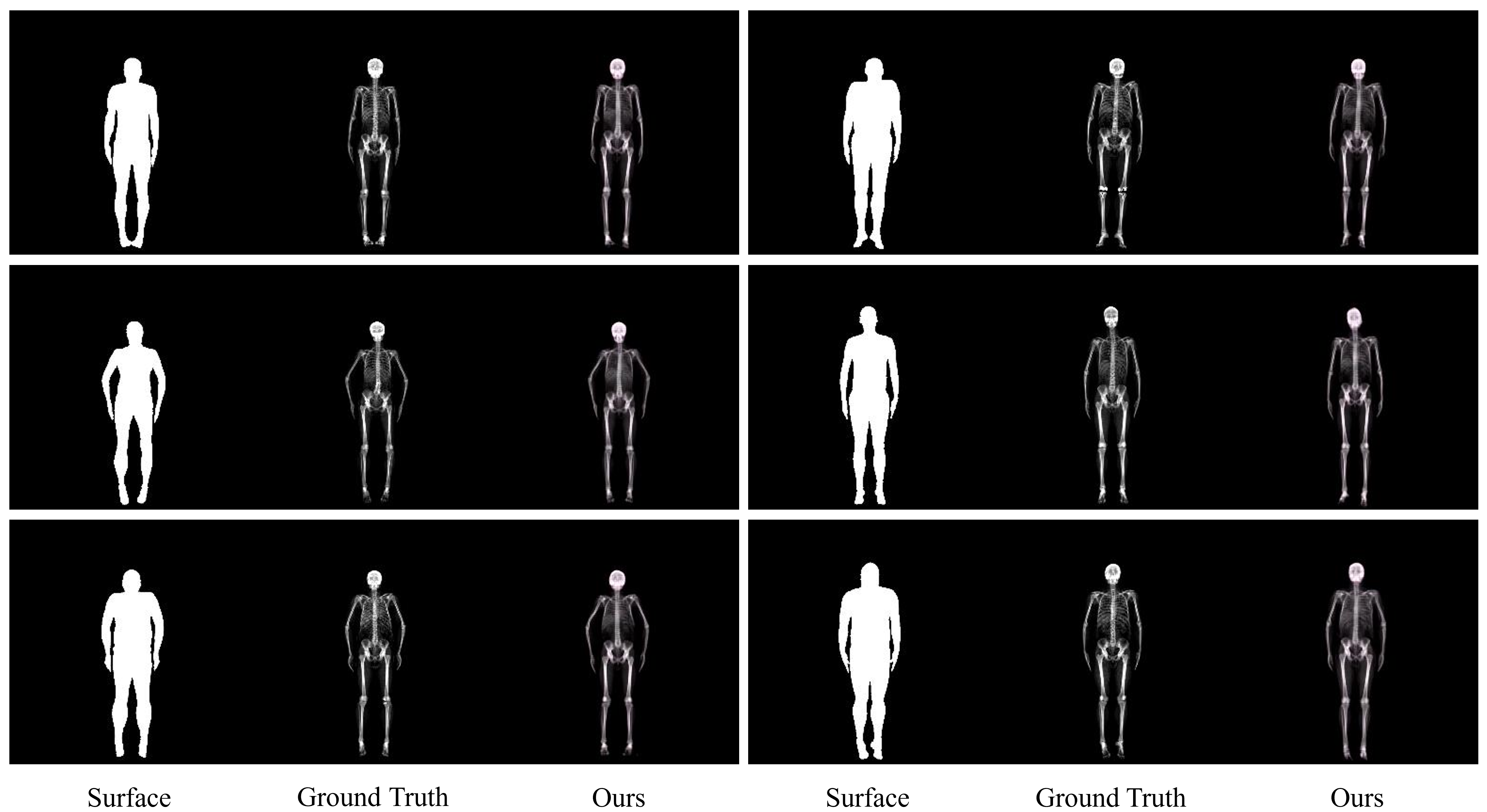}
  \caption{These pictures show the human masking image(left), Ground Truth X-ray image(middle), and generated X-ray image(right), from which we could see that our MaSkel could generate high-quality and well-structured human X-rays.}
  \label{img6}
\end{figure*}

\begin{table}
  \caption{Evaluation of Generated X-rays}
  \label{tab:freq}
  \begin{tabular}{cccc}
    \toprule
    Metrics&PSNR&SSIM&LPIPS\\
    \midrule
    $Ours_{R}$ & 33.82 & 0.9881 & 0.0210\\
    $Ours_{G}$ & 23.46 & 0.9206 & 0.0324\\
  \bottomrule
  \label{tabel1}
\end{tabular}
\end{table}

In Table\ref{tabel1}, $Ours_{R}$ indicates the results in the first-stage training, and $Ours_{G}$ shows the metric results in the second-stage training. The first line $Ours_{R}$ showed that our reconstructed data's PSNR is close to $34(dB)$. The SSIM is close to $1.0$, which could be thought that the reconstructed data is realistic enough. The LPIPS is close to $0.0$, which means the generated are similar well to the real one.

\subsection{The Second-stage Training}
In the second-stage training, our MaSkel generates the X-ray image from the masking one. The generated human skeleton X-ray images are compared with the ground truth images. Figure\ref{img6} shows that the generated results are quite similar to the ground truth, and they match the contours of the human body. We also evaluate the synthetic data by using PNSR, SSIM, and LPSPS in Table\ref{tabel1}. The $Ours_{G}$ indicates that PSNR is about $23.5(dB)$, the SSIM is $0.92$ and LPIPS is also close to $0$, which means our prediction has good reconstruction details and they are realistic enough compared to the ground truth.

\begin{figure}[h]
  \centering
  \includegraphics[width=\linewidth]{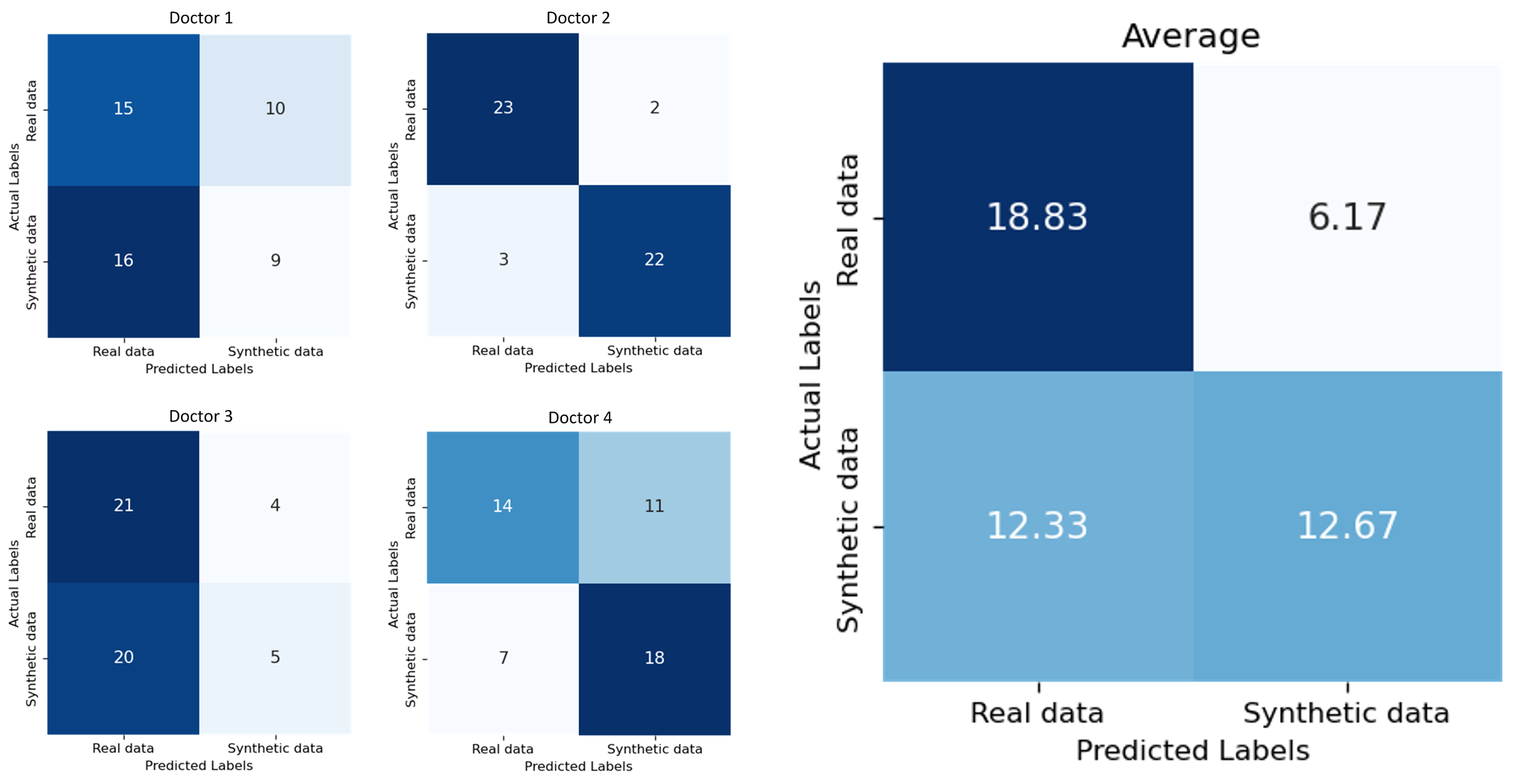}
  \caption{The confusion matrix reflects the doctor's evaluation of real X-ray images and generated ones. On the left is the evaluation display of four doctors, and on the right is the average evaluation of all doctors. The confusion matrix shows that our generated X-rays have somewhat confounded medical professionals, which indicates its high quality and credibility.}
  \label{img7}
\end{figure}

To verify the quality and authenticity of the X-ray images generated by our method, we further conducted a user study. We invited some professional doctors, tasking them to independently identify which of the 50 provided X-ray images were synthetic. Half of the X-ray images were generated using our method, while the remaining half consisted of authentic medical X-ray images. To guarantee the fairness of the evaluation, these images were randomized in advance. The doctors' identification results are quantified by constructing confusion matrices. Figure\ref{img7} displays the confusion matrices for 4 of the invited doctors, as well as the average confusion matrix derived from the assessments of all participating doctors. In this project, we focus more on the metric of false negatives, the number of generated images that are incorrectly identified as real images. The figure demonstrated that doctors have a higher accuracy in identifying real X-rays, as indicated by the higher TP values. However, the average confusion matrix (TN=18.83, FP=6.17, FN=12.33, TP=12.67) revealed that despite the participating doctors being experienced professionals, all of them misjudged synthetic X-ray images as real to some extent. Overall, the X-ray images generated by our method have somewhat confounded medical professionals, which indicates its high quality and credibility. Compared with real X-rays, they thought that the image from our method was equivalent to X-ray bone imaging in terms of overall contour, bone structure, spine vertebral body arrangement, joint alignment, and other details.

\subsection{Evaluation on the Real-World Data}
\begin{figure}[h]
  \centering
  \includegraphics[width=\linewidth]{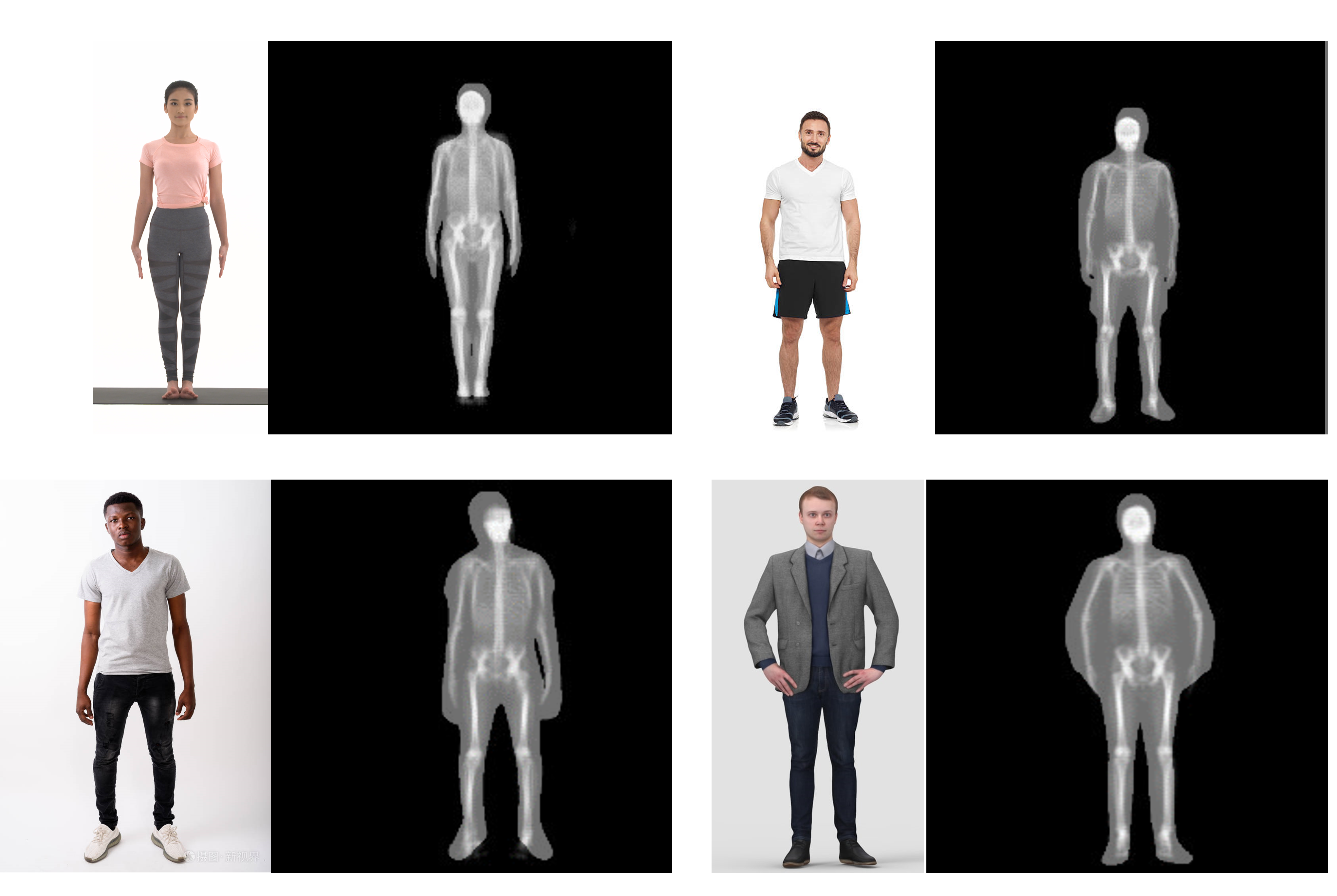}
  \caption{The visualization of real-world human X-ray generation. The image on the left is an image of a human body, and the image on the right is an overlapping image of a rough masking image and an X-ray image.}
  \label{img8}
\end{figure}

Our model has been trained on datasets comprising patients, characterized by their clear and fine contours. Our MaSkel demonstrates its exceptional capabilities in X-ray generation. However, to assess our model's adaptability and generalization across more varied and natural datasets, we utilize our model to predict the clothed human masking images from the real world. Without the CT scan, these images exhibit more coarse and irregular contours, significantly differing from our training data. We initially employ YOLO-v8\cite{reis2023real} to extract the human masking images from these wild images. Subsequently, this extracted surface image is processed to the special format the MaSkel model requires. Because these real-world images lack corresponding X-ray images as Ground Truth, thus our evaluation is primarily qualitative, focusing on the visual interpretation of the results. As indicated in Figure\ref{img8}, there is a noticeable degradation in reconstruction quality when applied to clothed individuals. This decline can be attributed to the less precise surface delineation due to clothing, which limits the ability of MaSkel. Despite this challenge, our MaSkel model still demonstrates a commendable level of resilience, delivering reasonable whole-body predictions that align roughly with the original surface images. This outcome highlights both the strengths and limitations of our model and points our research direction at the next stage.

\section{Conclusion}
We introduce a new approach, MaSkel is designed to generate human X-rays directly from the human masking image. To our knowledge, this innovative method is the inaugural technique to predict and assess 2D X-rays from human masking observations. Notably, on the data side, we have developed and released two synthetic datasets of human X-ray images we built with resolutions of $64 \times 64$ and $256 \times 256$, making some contributions to future research in this field.

In our work, we invite some professional doctors to do a serious user study. The feedback from a user study has confirmed that the X-ray images generated by our method closely resemble authentic X-ray photographs. Generated data includes essential medical details and almost accurate human anatomical structures. On the modeling side, MaSkel showcases exceptional predictive capabilities, particularly with images of non-clothed humans. The skeletal images produced not only maintain a high quality of generated data but also exhibit superior alignment with the surface images.

Despite these advances, our work confronts certain challenges. One significant problem is the limited posture of the training data, which is exclusively sourced from subjects in lying-down positions. This restricts the posture representational scope of our model. Additionally, our training dataset consists exclusively of finely paired images. Consequently, for images of clothed humans that present more irregular surface contours, there is a noticeable decline in the quality of the results produced by our model.

In our further work, we are planning to enhance our methodology by integrating MaSkel with SKEL, a sophisticated 3D skeleton parametric model. Our MaSkel could provide an accurate enough X-ray image as a reference, and SKEL provides a driveable 3D skeleton. In this way, this fusion could generate anatomically accurate human skeletons capable of depicting a broader array of actions, thereby overcoming current limitations and opening new avenues for research and application in medical imaging and beyond.

\newpage
\newpage

\begin{acks}
To Robert, for the bagels and explaining CMYK and color spaces.
\end{acks}

\bibliography{sample-base}

\appendix

\end{document}